\documentstyle[12pt]{article}

\begin{document}
\bibliographystyle{unsrt}

\vspace{1in}

\begin{center}
{\huge Noncommutative Geometry from String Theory:  Annulus Corrections}

\vspace{0.5in}

Mark Laidlaw \footnote{email: laidlaw@physics.ubc.ca}

\vspace{0.5in} 

University of British Columbia

6224 Agricultural Road 

Vancouver BC, Canada

V6T 1Z1

\vspace{0.5in}

{\large Abstract}
\end{center}

We develop a method in which it is possible to calculate
one loop corrections to the noncommutativity parameter
found for open strings in a background $F_{\mu\nu}$ field.
We first reproduce the well known disk results for $\theta^{\mu\nu}$.
We then consider the case of charged and neutral open strings
on the brane, and show in both cases that the result is the same
as in the disk case, apart from a multiplicative factor due to the
open string tachyon.
We also consider the case of open strings stretched between two branes
and show that our method reproduces known disk results.

\newpage

\section{Introduction}

In the past few years there has been much activity
exploring the stringy origins of non-commutative geometry.
\cite{gg,aa,qq,ss,vv}  Most of this activity has been done
by considering a bosonic string
in an antisymmetric background with the disk topology and
examining the corresponding commutators of the $X^\mu$ fields.
Broadly speaking there are two approaches in the literature; one
is to find a mode expansion for the $X$ fields which is orthogonal
and consistent with the boundary conditions induced by the background
field, and after calculating the commutator of the mode coefficients
using this to work out the full commutator between the $X$s 
\cite{tt,vv,ss,rr}, while
the other is to  calculate an exact propagator for the bosonic excitations
in the background and directly from that read off the commutator 
\cite{gg}.  Both of these methods are effective, and give the
same answer for the non-commutativity induced by the background field.
However, neither generalizes well to higher topologies, such as the annulus.
In the case of the operator approach, it relies on particular
assumptions of similar block diagonal structures for the gauge fields
on different branes, while exact propagators in the presence of 
an $F_{\mu\nu}$ background are complicated for higher genus topology.
\cite{ll,hh}.

With this in mind, we offer a slightly different way of deriving the
non-commutativity parameter for a bosonic string in an antisymmetric
background.  We show that this method of calculation reproduces the
known results for the disk, and proceed to calculate annulus 
level corrections to non-commutative geometry via an obvious generalization.
While many aspects of the annulus geometry are discussed in the
literature, for example \cite{xxxx, mm, uu},
this method produces new higher order
corrections.
The procedure is as follows: Since the Lagrangian for a string in a 
background field is 
\begin{equation}
S = \frac{-1}{4\pi\alpha'} \int d\sigma d\tau
\sqrt{-g} g_{ab} \partial^a X^\mu \partial^b X_\mu
+ \frac{1}{2} \int_{\partial \Sigma}
d\tau F_{\mu\nu} X^\nu \partial_\tau X^\mu |_{\sigma=a}
\end{equation}
we treat the interaction with  $F_{\mu\nu}$ as a perturbation, and
find the modification to the propagator due to $N$ interactions with
this background.  We then perform the sum over $N$ to obtain an exact
propagator and then using the prescription, due to Seiberg and Witten,
\cite{gg}, that
\begin{equation}
\left[ X^\mu(\tau),X^\nu(\tau) \right] = \lim_{\epsilon \rightarrow 0} {\cal T}
\left( X^\mu(\tau) X^\nu(\tau - \epsilon) 
- X^\mu(\tau) X^\nu(\tau +\epsilon) \right)
\label{eq:commdef}
\end{equation}
where ${\cal T}$ represents time ordering, we calculate the commutator,
and find a non-trivial relation.

\section{The Disk}

In the case of the disk it is well known that the 
free Green's function is
\cite{nn} 
\begin{equation}
G^{\mu\nu}
(z,z') = -\alpha'  \ln \left| z-z' \right| \left| z-\bar z'^{-1} \right|
\delta^{\mu\nu}. 
\label{eq:dfp}
\end{equation} 
Using the parameterization $z= \rho e^{i \phi}$, with $\rho \in \left(0,1
\right)$ and $\phi \in \left(0 , 2 \pi \right)$ the propagator can
be re-written on the boundary as
\begin{equation}
G^{\mu\nu}(e^{i \phi}, e^{i \phi'} )
= 2 \alpha' \sum_{m=1}^\infty \frac{ \cos m (\phi-\phi') }{m} \delta^{\mu\nu}. 
\label{eq:dep}
\end{equation}
Keeping in mind that the interaction term is
\begin{equation}
L_{int} = \frac{1}{2} \int d\phi F_{\mu\nu} X^\nu \partial_\phi X^\mu,
\end{equation}
it is clear that the contribution to the propagator due to $N$ interactions
with the background field is
\begin{eqnarray}
G^{\mu\nu}_N(z,z') &=& (F^N)^{\mu\nu} \int d\theta_1 \ldots
d\theta_N G(\rho e^{i \phi}, e^{i \theta_1}) \nonumber \\
&~& \times \partial_{\theta_1} 
G(e^{i \theta_1}, e^{i \theta_2}) \ldots \partial_{\theta_N} G(e^{i \theta_N},
\rho' e^{i\phi'} ).
\end{eqnarray}
Using (\ref{eq:dfp}) and (\ref{eq:dep}) in the above, and noting that the
integrals become trivial, we find that
\begin{equation}
G^{\mu\nu}_N(z,z') =
\alpha' \left( ( -2\pi \alpha' F )^N \right)^{\mu\nu}
\sum_{m=1}^\infty \frac{\rho^m}{m} (\rho'^m + \rho'^{-m} )
\left\{ \matrix{ \cos m (\phi-\phi')~N~even \cr
\sin m (\phi-\phi')~N~odd~ } \right\}. 
\end{equation}
So, summing over $N$, the full propagator taking into account 
the background field is
\begin{eqnarray}
G^{\mu\nu}(z,z') &=& \alpha' \left( \frac{1}{1 - (2 \pi \alpha' F)^2}
\right)^{\mu\nu} \sum_{m=1}^\infty \frac{ (\rho \rho')^m + (\rho/\rho')^m}{m}
\cos m (\phi-\phi') 
\nonumber \\
&~& + \alpha' \left( \frac{ -2 \pi \alpha' F}{1 - (2 \pi \alpha' F)^2}
\right)^{\mu\nu} \sum_{m=1}^\infty \frac{ (\rho \rho')^m + (\rho/\rho')^m}{m}
\sin m (\phi-\phi').
\nonumber \\
\end{eqnarray}
Now, applying this to (\ref{eq:commdef}) we see that
\begin{equation}
\left[ X^\mu, X^\nu \right] = \lim_{\epsilon \rightarrow 0}
2 \alpha' \left( \frac{ -2 \pi \alpha' F}{1 - (2 \pi \alpha' F)^2}
\right)^{\mu\nu} \sum_{m=1}^\infty \frac{ (\rho \rho')^m + (\rho/\rho')^m}{m}
\sin m \epsilon.
\end{equation}
In the case that $\rho$ and $\rho'$ are not both on the boundary (where
they would be equal to 1) the $\rho\rho'$ term 
converges to
something proportional to $\tan^{-1} \left( \frac{2 \rho \rho'
\sin\epsilon}{1-(\rho\rho')^2} \right)$,
which vanishes in the limit $\epsilon \rightarrow 0$, with something
analogous happening to the other term.  Thus, we see that the only
non-vanishing contribution comes from $\rho=\rho'=1$, and so in that
case
\begin{equation}
\left[ X^\mu, X^\nu \right] = \lim_{\epsilon \rightarrow 0}
2 \alpha' \left( \frac{ -2 \pi \alpha' F}{1 - (2 \pi \alpha' F)^2}
\right)^{\mu\nu} \sum_{m=1}^\infty \frac{ 2 \sin m \epsilon}{m}. 
\end{equation}
Now, using $2\sum_m \frac{\sin m x}{m} = \pi - x$, this gives
\begin{equation}
\left[ X^\mu, X^\nu \right] = 
2 \pi \alpha' \left( \frac{ -2 \pi \alpha' F}{1 - (2 \pi \alpha' F)^2}
\right)^{\mu\nu}
\end{equation}
as the commutator for the coordinates on the D-brane with a constant
$F_{\mu\nu}$ field on it, in agreement with previous results.

\section{The Annulus}
For the annulus, more care is needed, but we proceed in analogy with
the above development.  There are, in general, two situations that
must be considered, the first is the case where both ends of the
open string end on the same D-brane, in which case it provides a 
one loop correction to the noncommutativity parameter calculated 
previously, and the second is where each end is on a distinct D-brane.

The free 
Greens function for a boson on an annulus of inner radius $a$, and
circumference $2\pi$ can be shown to be
\begin{eqnarray}
G^{\mu\nu} (\rho e^{i\phi}, \rho' e^{i \phi'} ) &=& \frac{2 \alpha' }{\ln a}
\left( \sum_{m>0, n}  
\cos\frac{n \pi \ln \rho}{\ln a} \cos\frac{n \pi \ln \rho'}{\ln a} 
\frac{\cos m (\phi-\phi')
}{m^2 + \left( \frac{n \pi}{\ln a} \right)^2}
\right.  \nonumber \\
&~& \left.
+ 
\sum_{n>0} \cos\frac{n \pi \ln \rho}{\ln a} 
\cos\frac{n \pi \ln \rho'}{\ln a} \left( \frac{\ln a}{n \pi}\right)^2
\right) \delta^{\mu\nu}, 
\end{eqnarray}
which is equivalent under the identification $z = \rho e^{ i \phi}$ to the
Greens function for the annulus given in \cite{ll,oo}.

Now, using a technique that follows \cite{nn} 
we note that for $z$ and $z'$ both on the boundary of the annulus, the 
Greens function can be compactly expressed as
\begin{equation}
G^{\mu\nu} (\phi, \phi') = 2\alpha' \sum_{m=1}^\infty \frac{1}{m}
G_m \cos m (\phi - \phi') \delta^{\mu\nu},
\end{equation}
where
\begin{equation}
G_m = \left( \matrix{ A_m & B_m \cr B_m & A_m } \right),~A_m = 
\frac{1+a^{2m}}{1-a^{2m} },~B_m = \frac{2 a^m }{1-a^{2m} }.
\end{equation}
Similarly, the propagator between an arbitrary point on the annulus, $z$, and
a point on either edge, parameterized by $\phi$, can be given by 
the row vector
\begin{equation}
G^{\mu\nu}( \rho e^{i \phi}, \phi') = \frac{2 \alpha'}{\ln a}
\sum_{m=1}^\infty \sum_n 
\frac{1}{m^2 + \left( \frac{n \pi}{\ln a} \right)^2}
\left( \cos\frac{n \pi \ln \rho}{\ln a}, 
\cos\frac{n \pi \ln \rho}{\ln a}
(-1)^n \right) 
\delta^{\mu\nu}, 
\end{equation}
and interchanging the arguments gives the transpose of this.
Note that in this language, the interaction is given by
\begin{equation}
L_{int} = \frac{1}{2} \int d\phi \Omega_{\mu\nu} X^\nu \partial_\phi X^\mu,~
\Omega_{\mu\nu} = \left( \matrix{ F_{1\mu\nu} & 0 \cr 0 & F_{2\mu\nu} } 
\right),
\label{eq:ail}
\end{equation}
where $F_1$ and $F_2$ are the field strengths at the distinct ends of the 
brane.

We first consider the case of an annulus diagram for a string attached at both
ends to the same D-brane.  The interaction of a bosonic string with a background
field is equivalent to a Wilson loop inserted at its boundaries \cite{pp}  
and since the background is a constant field the orientation of the Wilson loop 
determines the sign of the charge at the endpoints of the string.  
There are apparently two inequivalent scenarios for this topology, the loops could
be identically or oppositely oriented, corresponding to the cases of charged and
neutral strings. 

\subsection{Annulus Corrections: Charged String}
The charged string corresponds to the case of identically oriented Wilson 
loops, and therefore, we have $F_1=F_2 = F$. 
It is also important to note that there are contributions to the commutator
from both ends of the string, and between opposite ends, since they all
end on the same brane.

As in the case of the disk, we calculate the correction to the propagator due
to $N$ interactions with the  background, and find it to be
\begin{eqnarray}
G^{\mu\nu}_N (\rho e^{i \phi},\rho' e^{i \phi'} ) &=&
(-2\alpha'\pi)^N \frac{2 \alpha'}{\ln^2 a} \left( F^N \right)^{\mu\nu}
\sum_{m=1}^\infty \frac{1}{m} \left\{ \matrix{ \sin m (\phi-\phi')~N~odd \cr
\cos m (\phi-\phi')~N~even } \right\}
\nonumber \\
&~& \times 
\sum_n 
\frac{1}{m^2 + \left( \frac{n \pi}{\ln a} \right)^2}
\left( \cos\frac{n \pi \ln \rho}{\ln a}, 
\cos\frac{n \pi \ln \rho}{\ln a}
(-1)^n  \right)
G_m^{N-1} 
\nonumber \\
&~& \times
\sum_{n'} \frac{1}{m^2 + \left( \frac{n'\pi}{\ln a} \right)^2}
\left( \matrix{ \cos\frac{n'\pi \ln \rho}{\ln a}
\cr
\cos\frac{n'\pi \ln \rho}{\ln a}
(-1)^{n'} } \right).
\label {eq:annchNcorr}
\end{eqnarray}
We find upon calculation that 
\begin{eqnarray}
G^{\mu\nu}_N (e^{i \phi},e^{i \phi'} ) &=& G^{\mu\nu}_N (a
e^{i \phi},ae^{i \phi'} )
\nonumber \\
&=& \alpha' (-2 \alpha' \pi)^N \left( F^N \right)^{\mu\nu}
\left( \frac{1}{1-a^{2m} } \right)^{N+1} 
\left[ (1+a^m)^{2N+2} \right. \nonumber \\
&~& \left. + (1-a^m)^{2N+2} \right]
\sum_{m=1}^\infty \frac{1}{m} \left\{ \matrix{ \sin m (\phi-\phi')~N~odd \cr
\cos m (\phi-\phi')~N~even } \right\},
\nonumber \\
\end{eqnarray}
while
\begin{eqnarray}
G^{\mu\nu}_N (e^{i \phi},a e^{i \phi'} ) &=& G^{\mu\nu}_N (a
e^{i \phi},e^{i \phi'} )
\nonumber \\
&=& \alpha' (-2 \alpha' \pi)^N \left( F^N \right)^{\mu\nu}
\left( \frac{1}{1-a^{2m} } \right)^{N+1} 
\left[ (1+a^m)^{2N+2} \right. \nonumber \\
&~& \left. - (1-a^m)^{2N+2} \right] 
\sum_{m=1}^\infty \frac{1}{m} \left\{ \matrix{ \sin m (\phi-\phi')~N~odd \cr
\cos m (\phi-\phi')~N~even } \right\}.
\nonumber \\
\end{eqnarray}

The obvious generalization of (\ref{eq:commdef}) is to sum the commutators
when the fields are at the same ends of the string with those where the fields
are at opposite ends of the string, since the ends are on the same brane.
In other words, to calculate 
\begin{eqnarray} 
\left[ X^\mu, X^\nu \right] &=& \frac{1}{4} \left(
\left[ X^\mu(1), X^\nu(1) \right] +\left[ X^\mu(1), X^\nu(a)\right]
\right. \nonumber \\
&~& \left. +\left[X^\mu(a), X^\nu(1) \right]+\left[X^\mu(a),X^\nu(a)\right]
\right).
\end{eqnarray} 
We thus find for a fixed value of $a$ that
\begin{eqnarray} 
\left[ X^\mu, X^\nu \right] &=& \sum_{N~odd} \sum_{m=1}^\infty
\lim_{\epsilon \rightarrow 0} (2\alpha')^{N+1} (-\pi)^N \left( F^N \right)^{\mu\nu}
\left( \frac{1+a^m}{1-a^m} \right)^{N+1} \frac{2}{m} \sin m \epsilon
\nonumber \\
&=& \sum_{m=1}^\infty \lim_{\epsilon \rightarrow 0}
2\alpha' \left( \frac{ -2 \pi \alpha' F k^2 }{ 1- (2 \pi \alpha' F k)^2 }
\right)^{\mu\nu} \frac{2}{m} \sin m \epsilon,
\end{eqnarray}
where $k= \frac{1+a^m}{1-a^m}$.

Now, to calculate the full commutator, it is necessary to integrate over the
Teichmuller parameter, with the measure as given in \cite{nn,cc}, so
\begin{eqnarray}
\left[ X^\mu, X^\nu \right] &=& 
\int_0^1 \frac{da}{a} 
\sum_{m=1}^\infty \lim_{\epsilon \rightarrow 0}
2\alpha' \left( \frac{ -2 \pi \alpha' F k^2 }{ 1- (2 \pi \alpha' F k)^2 }
\right)^{\mu\nu} \frac{2}{m} \sin m \epsilon.
\end{eqnarray}
We note that $F_{\mu\nu}$ can be block diagonalized, and concentrate
on one block: $X^{2i}, X^{2i+1}$.  After trivial manipulations,
we have
\begin{eqnarray}
\left[ X^{2i}, X^{2i+1} \right] &=& 
\int_0^1 \frac{da}{a} 
\sum_{m=1}^\infty \lim_{\epsilon \rightarrow 0}
2\alpha' \frac{2}{m} \sin m \epsilon
\frac{\beta}{1+\beta^2} \frac{1+2a^m +a^{2m} }{ 1+a^{2m}
- 2a^m \left( \frac{1-\beta^2}{1+\beta^2} \right) }
\nonumber \\
\end{eqnarray}
where $\beta = - 2\pi \alpha' F$, and $F$ is the field strength in this
particular block.
The $m$ dependence in the integral can be accommodated by the
change of variables, noting that $\frac{da}{a} = \frac{1}{m} \frac{d(a^m)}{
a^m}$
and this gives
\begin{eqnarray}
\left[ X^{2i}, X^{2i+1} \right] &=& 
\sum_{m=1}^\infty \lim_{\epsilon \rightarrow 0}
2\alpha' \frac{2}{m} \sin m \epsilon
\frac{\beta}{1+\beta^2} 
\nonumber \\
&~& \times
\frac{1}{m} \left[ \ln x |_{x=0}^{x=1} + \frac{2}{\beta} \tan^{-1} 
\beta + \frac{2}{\beta} \tan^{-1} \frac{1-\beta^2}{2\beta} \right]. 
\end{eqnarray}
Clearly, the only term that is divergent is the term $\ln x$, and if
we change the original lower limit of integration to $\delta \rightarrow 0$
and so change the lower limit on $x$ to $\delta^m$, 
we obtain for this particular block
\begin{eqnarray}
\left[ X^{2i}, X^{2i+1} \right] &=& 2 \pi \alpha' \frac{\beta}{1+\beta^2}
\lim_{\delta \rightarrow 0} \ln \frac{1}{\delta}, 
\end{eqnarray}
and since the other blocks will have the same structure, we find
\begin{eqnarray}
\left[ X^\mu,X^\nu \right] &=& 2 \pi \alpha' 
\left( \frac{ -2 \pi \alpha' F}{1 - (2 \pi \alpha' F)^2}
\right)^{\mu\nu}
\lim_{\delta \rightarrow 0} \ln \frac{1}{\delta}.
\end{eqnarray}
The coefficient of the divergent part is exactly the same as
in the case of the disk, and the divergence is interpreted 
\cite{xx} as coming from the open string tachyon.

\subsection{Annulus Corrections:  Neutral String}
The case where the two Wilson loops are oppositely oriented
gives the neutral string.  
Because of this in equation (\ref{eq:ail}) we have $F_1 =-F_2 = F$
and the analog of equation (\ref{eq:annchNcorr}) is 
\begin{eqnarray}
G^{\mu\nu}_N (\rho e^{i \phi},\rho' e^{i \phi'} ) &=&
(-2 \alpha' \pi)^N \frac{2 \alpha'}{\ln^2 a} \left( F^N \right)^{\mu\nu}
\sum_{m=1}^\infty \frac{1}{m} \left\{ \matrix{ \sin m (\phi-\phi')~N~odd \cr
\cos m (\phi-\phi')~N~even } \right\}
\nonumber \\
&~& \times
\sum_n \frac{1}{m^2 + \left( \frac{n \pi}{\ln a} \right)^2}
\left( \cos\frac{n \pi \ln \rho}{\ln a}, 
\cos\frac{n \pi \ln \rho}{\ln a}
(-1)^n \right)
\nonumber \\
&~& \times
\Omega 
(G_m \Omega)^{N-1}
\sum_{n'} \frac{1}{m^2 + \left( \frac{n'\pi}{\ln a} \right)^2}
\left( \matrix{ \cos\frac{n'\pi \ln \rho}{\ln a}
\cr
\cos\frac{n'\pi \ln \rho}{\ln a}
(-1)^{n'} } \right).
\end{eqnarray}
In the above, $\Omega = \left( \matrix{ 1 & 0 \cr 0 & -1} \right)$, which
is the field strength independent part of the interaction vertex.
Upon diagonalizing $G_m \Omega$ we find that
\begin{eqnarray}
G^{\mu\nu}_N (\rho e^{i \phi},\rho' e^{i \phi'} ) &=&
(-2 \alpha' \pi)^N \frac{2 \alpha'}{\ln^2 a} \left( F^N \right)^{\mu\nu}
\sum_{m=1}^\infty \frac{1}{m} \left\{ \matrix{ \sin m (\phi-\phi')~N~odd \cr
\cos m (\phi-\phi')~N~even } \right\}
\nonumber \\
&~& \times
\sum_n \frac{1}{m^2 + \left( \frac{n \pi}{\ln a} \right)^2}
\left( \cos\frac{n \pi \ln \rho}{\ln a}, 
\cos\frac{n \pi \ln \rho}{\ln a}
(-1)^n \right)
\nonumber \\
&~&\times
\left( \matrix{ 1 & 0 \cr 0 & -1} \right)
\left( \matrix{ 1 & -a^m \cr a^m & -1 } \right)
\left( \matrix{ 1 & 0 \cr 0 & (-1)^{N-1} }\right) 
\left( \matrix{ 1 & -a^m \cr a^m & -1 } \right)
\nonumber \\
&~&\times
\left( \frac{1}{1-a^{2m} } \right)
\sum_{n'} \frac{1}{m^2 + \left( \frac{n'\pi}{\ln a} \right)^2}
\left( \matrix{ \cos\frac{n'\pi \ln \rho}{\ln a}
\cr
\cos\frac{n'\pi \ln \rho}{\ln a}
(-1)^{n'} } \right).
\end{eqnarray}

As in the case of the charged string, since the ends of the string
terminate on the same brane, it is necessary to sum the contribution
of all possible commutators to determine the total commutator.
However, the above equation has a structure that is easy to analyze:
the commutator of crossover terms, like $\left[ X^{\mu}(1), X^\nu(a)
\right]$ explicitly vanish, and $\left[ X^\mu(1), X^\nu(1) \right] = 
- \left[ X^\mu(a),
X^\nu (a) \right] $, so the annulus level commutator for a neutral string
explicitly vanishes.
In retrospect, this is not a surprising result.  As shown above, for
a charged bosonic string the commutation relation at annulus level 
consists of a constant, identical to that for the disk level calculation
multiplied by the tachyon divergence.  
It is trivial to see that for an uncharged string there is no 
disk level contribution to the nontrivial commutator because the uncharged
string does not couple to the background field, and we show that this
property persists at one loop.

\subsection{Annulus Corrections:  Distinct Branes}
We now examine the most general case, that of two distinct branes
with independent background fields on each.  Following the 
analysis of the previous sections, we find that
the correction due to $N$ interactions with the backgrounds
is
\begin{eqnarray}
G^{\mu\nu}_N (\rho e^{i \phi},\rho' e^{i \phi'} ) &=&
(-2 \alpha' \pi)^N \frac{2 \alpha'}{\ln^2 a}
\sum_{m=1}^\infty \frac{1}{m} \left\{ \matrix{ \sin m (\phi-\phi')~N~odd \cr
\cos m (\phi-\phi')~N~even } \right\}
\nonumber \\
&~& \times
\sum_n \frac{1}{m^2 + \left( \frac{n \pi}{\ln a} \right)^2}
\left( \cos\frac{n \pi \ln \rho}{\ln a}, 
\cos\frac{n \pi \ln \rho}{\ln a}
(-1)^n \right)
\nonumber \\
&~& \times
\Omega^{\mu\alpha_1} 
G_m \ldots G_m \Omega^{\alpha_{N-1} \nu}
\sum_{n'} 
\left( \matrix{ \cos\frac{n'\pi \ln \rho}{\ln a}
\frac{1}{m^2 + \left( \frac{n'\pi}{\ln a} \right)^2}
\cr
\cos\frac{n'\pi \ln \rho}{\ln a}
\frac{(-1)^{n'}}{m^2 + \left( \frac{n'\pi}{\ln a} \right)^2}
 } \right).
\end{eqnarray}
Since the two branes are distinct, it is only sensible to consider the
commutator between $X$ fields at the same end 
of the string.
Explicitly we find that
\begin{eqnarray}
G^{\mu\nu}_N(  e^{i \phi},  e^{i \phi'} ) &=&
2\alpha' (-2 \alpha' \pi)^N  
\sum_{m=1}^\infty \frac{1}{m} \left\{ \matrix{ \sin m (\phi-\phi')~N~odd \cr
\cos m (\phi-\phi')~N~even } \right\}
\nonumber \\
&~& \times
\sum_n \left( coth m\ln a, csch m\ln a \right) 
\Omega^{\mu\alpha_1}
G_m \ldots \nonumber \\
&~& \times
G_m \Omega^{\alpha_{N-1} \nu}
\left( \matrix{ coth m\ln a \cr csch m\ln a } \right).
\end{eqnarray}
We consider this in the limit of $a \rightarrow 0$, which corresponds
to the disk amplitude.  
In this limit, we have
\begin{eqnarray}
G^{\mu\nu}_N(  e^{i \phi},  e^{i \phi'} ) &=&
(2\alpha')^{N+1} (-\pi)^N (F_1^N)^{\mu\nu} 
\sum_{m=1}^\infty \frac{1}{m} \left\{ \matrix{ \sin m (\phi-\phi')~N~odd \cr
\cos m (\phi-\phi')~N~even } \right\}, 
\nonumber \\
\end{eqnarray}
so on the first brane, 
\begin{equation}
\left[ X^\mu, X^\nu \right] = 2\pi \alpha' \left( 
\frac{ -2\pi \alpha' F_1}{
1-(2\pi\alpha' F_1)^2} \right)^{\mu\nu}.
\end{equation}
Similarly, it is clear that in this limit on the second brane we
have
\begin{equation}
\left[ X^\mu, X^\nu \right] = 2\pi \alpha' \left( 
\frac{ -2\pi \alpha' F_2}{
1-(2\pi\alpha' F_21)^2} \right)^{\mu\nu}.
\end{equation}
The cross terms are clearly zero.  This result reproduces and
generalizes that found in \cite{tt} for noncommutativity
on the ends of branes.  It is important to note that 
this result does not rely on any relations between the two
fields on the branes, in particular, they do not need to have
the same block diagonal form.

\section{Conclusion}
In this paper we have shown a method for calculating the
disk and annulus contributions to noncommutative geometry.
In particular, we have shown that the contribution from the
higher order annulus diagrams is that which comes from the
disk, multiplied by the tachyon divergence.  
The benefits of the method presented are that it does not depend
upon any particular properties of the background, other than
the fact it is constant.

\section{Acknowledgments}
This work is supported in part
by the Canadian Natural Sciences and Engineering
Research Council.

\end{document}